\documentclass[12pt]{article}
\usepackage{epsfig}
\newcommand{\nn}{\nonumber}

\newcommand{\be}{\begin{equation}}
\newcommand{\ee}{\end{equation}}
\newcommand{\bea}{\begin{eqnarray}}
\newcommand{\eea}{\end{eqnarray}}

\newcommand{\xm}{\left(1-2\gamma^2z\,(1-\beta\cos{\theta}^{'})\right)}
\newcommand{\x}{\left(2\gamma^2z\,(1-\beta\cos{\theta}^{'})\right)}
\newcommand{\J}{(1-\beta\cos{\theta}^{'})}

\begin{document}
%
\thispagestyle{empty}
\begin{flushright}
{\tt hep-ph/0203052}\\
{FTUAM-02-06}\\
{IFT-UAM/CSIC-02-04} \\

\end{flushright}
\vspace*{1cm}
\begin{center}
{\Large{\bf Corrections to the fluxes of a Neutrino Factory} }\\
\vspace{.5cm}
A. Broncano$^{\rm a,}$\footnote{alicia@delta.ft.uam.es},
O. Mena$^{\rm a,}$\footnote{mena@delta.ft.uam.es}
 
\vspace*{1cm}
$^{\rm a}$ Dept. de F\'{\i}sica Te\'orica, Univ. Aut\'onoma de
Madrid, 28049 Spain \\
\end{center}
\vspace{.3cm}
\begin{abstract}
\noindent
In view of their physics goals, future neutrino factories from muon
 decay aim at an overall flux precision of  ${\cal O}(1\%)$ or
 better. We analytically study the QED radiative corrections to the
 neutrino differential distributions from muon decay. Kinematic uncertainties due to the divergence of the muon beam are considered as well. The resulting corrections to the neutrino flux turn out to be
 of order ${\cal O}(0.1\%)$, safely below the required precision. 
\end{abstract}

\newpage
\pagestyle{plain} 
\setcounter{page}{1}
\setcounter{footnote}{0}
%
\section{Introduction}
Results on neutrino oscillations from Superkamiokande \cite{SuperK} and SNO \cite{SNO} provide a compelling evidence for neutrino 
masses, constituting the first  strong indication of Physics beyond the Standard Model. Much is still unknown, though, regarding fundamental 
issues such as 
the absolute neutrino mass scale, the possible Majorana character of neutrino fields, the ordering of their mass eigenstates with respect 
to charged lepton eigenstates, or the possible existence of leptonic 
CP violation and its tantalizing relationship to baryogenesis. In this situation one could argue that the subject of lepton flavor 
physics is at its exciting infancy, and to obtain rough answers to those questions could be a sufficient goal at present, 
postponing any aim at a precise determination of the involved parameters.
Nevertheless, some of those questions prerequire precision: for instance the study of CP violation rests upon a precise knowledge 
of the angles in the neutrino mixing matrix. 

In a more general way and much as for the quark sector, it is necessary to know accurately the values of the 
masses and mixing parameters in the lepton sector, as a first step to unravel the flavor puzzle. 
And what does precision means, quantitatively?. For instance, with which precision is it desirable to determine the values of the leptonic 
mixing angles in order to discriminate between models for neutrino masses? Clearly no definite answer can be given to such question, but as an 
indication it has been argued \cite{Snowmass} that a $10\%-1\%$ precision in the knowledge of, say, $sin^2 2\theta_{atm}$ would result 
in significant advance\footnote{$\theta_{atm}$ denotes the mixing angle dominantly responsible for the atmospheric oscillations, denoted by 
$\theta_{23}$ in the by now standard parameterization \cite{Particle}}. It is not impossible to envisage such a precision.
In resume, we are simultaneously entering 
a discovery and a precision era in neutrino physics. With the bonus that the extraction of physical conclusions will not be necessarily hindered by large 
theoretical errors, as it happens in the quark sector due to QCD long distance contributions.

 A quest for precise physics answers evidently requires an effort in precision on the experimental conditions, and on the knowledge of the 
neutrino flux to start with.
Several experiments using neutrino beams from particle accelerators such as K2K, MINOS and OPERA \cite{exper} will take data in the next few years.  
Their reach will be limited by the use of
 conventional neutrino beams produced from a charged pion source. The decay $\pi^+\rightarrow \mu^+\nu_\mu\, (\pi^-\rightarrow \mu^-\bar\nu_\mu)$ 
produces a  $\nu_\mu$ beam with a ${\cal O}(1\%)$ component of $\nu_e$ from kaon decays. The $\nu_e$ contamination limits the precision of 
the flux measurements, resulting in an error of $7\%$ for K2K, while MINOS reduces it to $2\%$ \cite{exper}.
A further step forward could be provided by the so-called superbeams which, although based on the same traditional beams, can achieve 
better precision thanks to the much higher statistics. It has been argued, for example, that by working at energies below the threshold of kaon production, the $\nu_e$ flavor 
contamination could be reduced, with the overall figure of merit for precision in the flux measurements limited to  ${\cal O}(1\%)$ 
\cite{todos,pilar}.

A major advance should come from a neutrino factory from muon decays,
aiming at both fundamental discoveries and ${\cal O}(1\%)$ precision
measurements. Present projects consider the production of very intense
muon sources of about $10^{20}$ muons per year \cite{Geer1}. Neutrino
beams originate from the decay of high-momentum muons along the straight
sections of a storage ring. The beam produced presents a precisely known
neutrino content: $50\%$ muon neutrinos and $50\%$ electron antineutrinos
if a $\mu^-$ beam is used, and $50\%$ muon antineutrinos and $50\%$
electron neutrinos if a $\mu^+$ beam is used. The resulting $\nu$ fluxes
are expected to be known with a precision better than $1\%$
\cite{Blondel}. It is necessary to ensure that any possible corrections
and sources of errors are controlled at that level.  In this work, we
study two effects:  the contribution of QED one-loop corrections to muon
decay and the divergence of the muon beam.  For both cases, we give novel
corrected formulae for neutrino differential distributions.

Radiative corrections to the electron differential distribution in
$\mu^{-} \rightarrow e^{-} + \bar\nu_e + \nu_\mu $ were calculated long
ago resulting in a correction of ${\cal  O}(1\%)$ \cite{Beherends},
larger than the expected effect of ${\cal  O}(\frac{\alpha}{\pi})\sim
{\cal O}(0.1\%)$. Such an effect is at the level of the expected
precision at a neutrino factory. In this work we study whether QED
corrections affect neutrino distributions at the same order.

The correction to the (massive) neutrino spectra from unpolarized muons has
been first calculated in \cite{Greub}. In our work, we give new anlytic
formulae with $m_\nu=0$ and $m_e=0$ including muon polarization,
relevant for neutrino factory measurements. Different from the
electron case, the analysis of the correction to the neutrino
differential distributions entails non-trivial integrations. By
using the correspondence between the QED corrections to the
$\mu$-decay and those for the charge $2/3$ heavy quarks in QCD, we 
make use of the techniques developped in
Refs.~\cite{Jezabek1,Jezabek2,Jezabek3} for the calculation of the QCD
corrections to the lepton spectrum in the decay $t\rightarrow b+l^++\nu_l$. 

The second subject addressed in this paper is that of the muon beam
divergence, one of the basic properties that can bias the predicted
neutrino spectra . We explore the error induced in the neutrino
distributions at the far site due to the systematic uncertainty on the
angular divergence, and compare our results with previous ones in which
this effect was not included \cite{golden}.

The paper is organized as follows. In section 2 we recall the tree-level
angular distributions. In section 3 the neutrino one-loop corrected
formulae are given, with 3.1 specializing in the soft photon
limit and cancellation of infrared divergences. Section 4 accounts for the
corrections due to the beam divergence.

\section{General definitions}
 
In the muon rest-frame, the angular distributions of the neutrinos produced in 
the decay $\mu^- \rightarrow e^- + \nu_\mu  + \bar\nu_e$, Fig.~\ref{loops}a, are computed from the muon decay rate:
\be
d\Gamma_0=\frac{1}{2m_{\mu}}\,64\,{\rm G_F}^2\,\,|M_0(p_\mu;p_e,p_{\bar\nu_e},p_{\nu_\mu}) |^2\,d\Phi_3(p_\mu;p_e,p_{\bar\nu_e},p_{\nu_\mu})~,
\ee
where $|M_0(p_\mu,p_e,p_{\bar\nu_e},p_{\nu_\mu}) |^2$ is the averaged squared amplitude obtained from the Feynmann diagram at tree level. 
For polarized muons:
\be
\label{tree_amp}
|M_0(p_\mu;p_e,p_{\bar\nu_e},p_{\nu_\mu}) |^2 = \left[(p_\mu - m_\mu s)\,p_{\bar\nu_e}\right](p_e p_{\nu_\mu})~,
\ee 
where $s$ is the four-spin. For unpolarized muons $s=0$.

$d\Phi_3$ is the three-body phase-space. In general, the n-body phase space is defined by :
\be
d\Phi_n(P;p_1,...,p_n) = (2\pi)^4\,\delta (P-p_1-...-p_n)
\,\prod_{i=1}^n
\frac{d^3{\bf p_i}}{2p_i^0}\frac{1}{(2\pi)^3}~. 
\ee

Differential distributions of decay products are obtained integrating over the phase space of the remaining decay particles,  
\bea 
\frac {d^2 N }{ dx\,d\cos{\theta} } = 
F^{(0)} (x) + J^{(0)}(x)\,{\cal P}_\mu \cos\theta~, 
\label{treedistr} 
\eea
where $x$ denotes the scaled energy, $x=2E_{e,\nu}/m_{\mu}$ and 
${\cal P}_\mu$ is the average over polarization  of the initial state muon along the beam direction.
$\theta$ is the angle between three-momentum of the emitted particle and 
the muon spin direction and $m_{\mu}$ is the muon mass. 
The normalized functions $F^{(0)}$ and $J^{(0)}$, in the limit $m_e=0$, read \cite{gaisser}:
\bea
F_{e}^{(0)}(x) &=&  x^2(3-2x)~,
\qquad
J_{e}^{(0)}(x) = x^2(1-2x)~,
\label{tree_e}\\
F_{\nu_\mu}^{(0)}(x) &=&  x^2(3-2x)~,
\qquad
J_{\nu_\mu}^{(0)}(x) = x^2(1-2x)~,
\label{tree_numu}\\
F_{\bar\nu_e}^{(0)}(x) &=& 6x^2(1-x)~,
\qquad
J_{\bar\nu_e}^{(0)}(x) = 6x^2(1-x)~.
\label{tree_nue}
\eea

\section{QED corrections}

The QED radiative corrections to the formula (\ref{tree_e}) where
calculated long ago in Ref.~\cite{Beherends} through the integration
over the neutrino phase-space of  the $\cal{O}(\alpha)$ corrected
differential muon decay rate. The QED corrections to the
Eq.~(\ref{tree_numu}) ( Eq.~(\ref{tree_nue}) ), are similarly obtained 
from the integration over the $\bar\nu_e-e^-$ ($\nu_\mu-e^-$) phase-space.  

In the muon decay, the QED corrected differential rate is given by
\be
{\rm d}\Gamma= {\rm d}\Gamma_0+{\rm d}\Gamma_V+{\rm d}\Gamma_R~,
\label{Gamma}
\ee 
where ${\rm d}\Gamma_V$ describes the contribution of virtual photon 
diagrams in Figs.~\ref{loops}b-\ref{loops}d and ${\rm d}\Gamma_R$
accounts for the effects of the real photon emission diagrams in   
 Fig.~\ref{loops}e and Fig.~\ref{loops}f.

The virtual photon  correction to the decay rate is given by
\be
\label{GammaV}
d\Gamma_V=\frac{1}{2m_{\mu}}\,64\,{\rm G_F}^2\,\,|M_V|^2\,d\Phi_3(p_\mu;p_e,p_{\bar\nu_e},p_{\nu_\mu})~,
\ee
where $|M_V|^2$ is the squared amplitude. For unpolarized muons it has
the expression:
\bea
\label{Mvirtual}
|M_V|^2= |M_0|^2\, -\,\frac{\alpha}{\pi}&\bigg[&
g_{\rm{L}}^{\rm{S}}\,|M_0|^2 
+\frac{m_\mu m_e}{4}\,g_{\rm{R}}^{\rm{S}}\,(p_{\bar\nu_e}p_{\nu_\mu})
\nn \\
&+& m_e\,g_{\rm{L}}^{\rm{V}}\,(p_\mu p_{\bar\nu_e})(p_\mu p_{\nu_\mu})
+m_\mu\,g_{\rm{R}}^{\rm{V}}\,(p_e p_{\bar\nu_e})(p_e p_{\nu_\mu})\,\,\bigg]~.
\eea
where $g_{\,\rm{L,R}}^{\rm{S,V}}$ are ultraviolet (UV) and are listed
in the appendix A. The function $g_{\rm{L}}^S$ is infrared (IR) divergent
while the rest of the ``$g$'' functions are finite. 

The IR singularity in  $g_{\rm{L}}^S$  is  canceled
with the soft photon terms of the real emission diagrams, which
correct the differential rate as  
\be
\label{GammaR}
d\Gamma_R=\frac{1}{2m_{\mu}}\,64\,{\rm G_F}^2\,\,|M_R|^2\,d\Phi_4(p_\mu;p_e,p_{\bar\nu_e},p_{\nu_\mu},k)~.
\ee
The explicit expression of the amplitude $|M_R|^2$ can be found in
the Appendix A.

QED corrections for polarized muons are calculated identically to
those of unpolarized ones with the replacement $p_\mu\rightarrow
p_\mu-sm_{\mu}$ in the amplitudes , where $s$ is the
muon four-spin \cite{Jezabek2}.  
 
\subsection{Soft photon limit and IR cancellation}

Before continuing with the discussion of the exact corrections, let us
consider their soft photon limit, i.e. $k \rightarrow 0$, as only IR 
singular terms of virtual and real photon diagrams remain in this
limit. 
When soft virtual and soft real photon contributions are added up, all ${\cal O}(\alpha)$ IR singularities are canceled.

In this limit, the ${\cal O}(k)$ terms in the virtual photon diagrams are neglected and Eq.~(\ref{Mvirtual}) is simplified to :
\be
\label{soft_Mvirtual}
|M_V^{\rm S P}|^2=  \left(1-\,\frac{\alpha}{\pi} \,  g_{\rm{L}}^{\rm{S}}\right)\,|M_0|^2, 
\ee
where $g_{\rm{L}}^{\rm{S}}$ contains all IR divergent terms which are
regularized introducing a finite photon mass
$\lambda$.

In the diagrams containing real photon emission, only terms of order
${\cal O}(k^{-2})$ remain in the soft photon limit. They contain all
IR divergent contributions from bremsstrahlung. The squared amplitude
in Eq.(\ref{GammaR}) reduces to: 
\be
\label{soft_ampl_real}
|M_R^{\rm S P}|^2= 32\,\frac{\alpha}{2\pi}\,\bigg[\frac{p_\mu^2}{(p_\mu k)^2}+\frac{p_e^2}{(p_e k)^2}-\frac{2(p_\mu p_e)}{(p_\mu k)(p_e k)}\bigg]|M_0|^2.
\label{IRreal}
\ee

The divergences in Eq.~(\ref{soft_Mvirtual}) cancel when added with
the soft bremsstrahlung part. However, Eq.~(\ref{soft_ampl_real}) must
be previously integrated over the photon-electron phase space in order
to reduce the real phton emission from a four-body problem to a three-body 
problem. The integral is performed introducing a finite photon mass
$\lambda$, resulting in a expression which exactly cancels the IR
terms in Eq.~(\ref{soft_Mvirtual}). 

After the integration over the corresponding phase space, we obtain the
soft-photon corrected for both $\nu_\mu$ and $\bar\nu_e$
distributions, which are proportional to the tree level
amplitude , in the limits $x\rightarrow 1$ and $m_e\rightarrow 0$:
\bea
\label{soft_dist}
\frac {d^2 N_{\nu}^{S P} }{ dx\,d\cos{\theta} } =   F_{\nu}^{(0)} (x)[1 - \frac{\alpha}{2\pi}\,k(x)].
\eea
The resulting function $k(x)$ is $\lambda$-independent:
\be
k(x) = 2\,{\rm L}(x) + 2 \pi^2/3 + \ln^2(1-x).
\ee
where ${\rm L}(x)$ is the Spence function defined in the Appendix A.

Realize that $k(x)$ diverges for $x\rightarrow 1$. This singularity is
originated due to a failure the perturbative treatment: at the end
point of the spectrum, the phase space for the emission of real
photons shrinks to zero and does not compensate the IR infinities of
the virtual photons.  

The end-point singularity of the corrected electron distribution from
muon decay has been largely discussed in the
literature \cite{Matson}. For $x \rightarrow 1$ and $k \rightarrow
0$, the corrected electron differential distribution diverges as
$\ln(1-x)$.  Since the IR divergences in the muon decay stem
from soft-photons in the limit $k\rightarrow 0$, the solution proposed 
to control the end-point divergence is to consider multiple
soft-photon emission. The effect of considering soft photons at
all orders in $\alpha$ is the exponentiation of the
singular logarithm $\ln(1-x)$ which leads to a non-singular distribution
\cite{Yennie}. 

Following as a guideline the solution found for the
electron, we apply the same proccedure to the neutrino
distributions. Consider the neutrino soft-photon correction in Eq.~(\ref{soft_dist}).  At the end-point, for each soft virtual photon and each soft real photon we get a $\ln^2(1-x)$ term, which multiplies the tree level amplitude. 
If there are $n$ soft virtual photons and $n$ soft real photons, there are $n$ double logarithms with an additional symmetry factor of $1/n!$. 
Therefore, the correction to the neutrino distribution at all orders in $\alpha$ is obtained summing over $n$:
\be
\frac {d^2 N_{\nu} }{ dx\,d\cos{\theta} } =   [F_{\nu}^{(0)} (x) + J_{\nu}^{(0)}\, {\cal P}_\mu(x) \cos \theta]\,e^{-\frac{\alpha}{2\pi}\ln^2(1-x)}~.
\ee

The evaluation of infrared divergences at all orders results in the
exponentiation of the double logarithm, which ensures a non-divergent
behavior of the neutrino distributions. The exponentiation is only valid for a small region $x\rightarrow 1$. For lower $x$, we 
must include all the terms of the exact corrections, computed in the next subsection.

\subsection{Results}

Exactly corrected neutrino distributions are obtained considering all 
terms of the ${\cal O}(\alpha)$ corrected decay rate,
Eq.~(\ref{Gamma}) and integrating over the phase space of the
remaining particles. Different from the corrected electron
distribution, in the neutrino case, the integrals over the
electron-photon phase space in the real emission diagrams are
nontrivial. We follow the method found in Ref.~\cite{Jezabek3} to
solve analytically these integrals in the calculation of the QCD
corrections to the lepton spectrum from the decay
$t\rightarrow b+l^++\nu_l$. We use the fact that there is a one to one correspondence
between the Feynmann diagrams in Fig.~\ref{loops} for the QED
corrections to the $\mu$-decay and those for the top
quarks. This correspondence can be seen by simply replacing 
\bea
\alpha&\to&\frac{4}{3}\alpha_S\nn\\
(\mu^-,e^-,\bar\nu_e,\nu_\mu)&\to&(t,b,l^+,\nu_l)~.
\eea

Therefore, by following the techniques detailed Ref.~\cite{Jezabek3}, we
perform the corresponding phase-space integrals to the differential
rate of polarised muons. We find that the  
corrected neutrino angular and energy distributions, including all finite terms in the limit $m_e=0$, are: 
\bea
\frac {d^2 N_{\nu_{\mu}} }{ dx\,d\cos{\theta} } =  
F_{\nu_\mu}^{(0)} (x) + J_{\nu_\mu}^{(0)}\, {\cal P}_\mu(x) \cos\theta 
- \frac{\alpha}{2\pi} \left[F_{\nu_{\mu}}^{(1)}(x)+  
J_{\nu_\mu}^{(1)}(x)\,{\cal P}_\mu \cos \theta\right]\,\,,\nn \\ \nn\\
\frac {d^2 N_{\bar\nu_e}}{ dx\,d\cos{\theta} } =    
F_{\bar\nu_e}^{(0)} (x) + J_{\bar\nu_e}^{(0)}(x)\,{\cal P}_\mu\cos\theta 
- \frac{\alpha}{2\pi} \left[F_{\bar\nu_e}^{(1)}(x) 
+ J_{\bar\nu_e}^{(1)}(x)\,{\cal P}_\mu\cos \theta\right]\,\,\\ \nn  \eea
where $F_{{\bar\nu_e},\nu_\mu}^{(0)}$-$J_{{\bar\nu_e},\nu_\mu}^{(0)}$ are given in Eq.~(\ref{tree_numu}), 
and the one-loop corrections are given by:
\bea
\label{F1mu}
F_{\nu_{\mu}}^{(1)}(x) & = &  F_{\nu_\mu}^{(0)}(x) k(x)
                 + \frac{1}{6}(41-36x+42x^2-16x^3)\ln(1-x)\nn \\
                 & + & \frac{1}{12}x(82-153x+86x^2)\,\,,  \\
\label{J1mu}
J_{\nu_{\mu}}^{(1)}(x) & = &  J_{\nu_\mu}^{(0)}(x) k(x)
                 + \frac{1}{6}(11-36x+14x^2-16x^3-4/x)\ln(1-x)\nn \\
                 & + & \frac{1}{12}(-8+18x-103x^2+78x^3)\,\,,  \\
\label{F1e}
F_{\bar\nu_e}^{(1)}(x)     & = & F_{\bar\nu_e}^{(0)}(x) k(x) 
                 +  (1-x)\Big[(5+8x+8x^2)\ln(1-x)\nn \\
                 & + & \frac{1}{2}x(10-19x)\Big] \,\,,\\\nn \\
\label{J1e}
J_{\bar\nu_e}^{(1)}(x)     & = & J_{\bar\nu_e}^{(0)}(x)k(x)
                 + (1-x)\Big[(-3+12x+8x^2+4/x)\ln(1-x) \nn \\
                 & + & \frac{1}{2}(8-2x-15x^2)\Big]\,\,.
\eea

As expected, due to the above correspondence, the results in
Eqs.~(\ref{F1mu})-(\ref{J1e}) are identical to those for the QCD corrections of the lepton distributions from top decay. 

Notice that the function
$k(x)$ appears in Eqs.~(\ref{F1mu})-(\ref{J1e}) multiplying the tree level
functions $F_{\nu}^{(0)}$-$J_{\nu}^{(0)}$, which agrees with the
discussion in the former subsection.

\begin{figure}[ht]
\centering
\epsfig{file=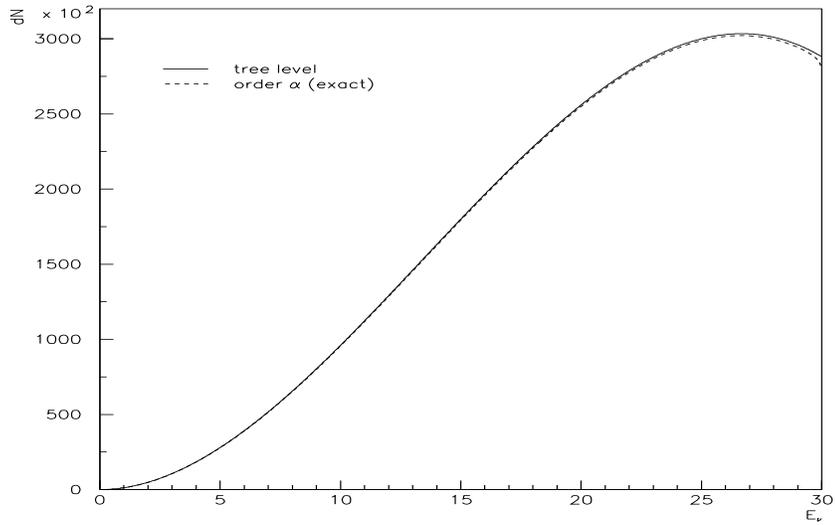, height=8cm,width=12cm}
\caption{\footnotesize Zeroth order and ${\cal O}(\alpha)$ corrected
$\nu_\mu$ forward distributions for $E_\mu=30$ GeV and ${\cal P}_\mu$=0.2.}
\label{numu}
\end{figure}

\begin{figure}[ht]
\centering
\epsfig{file=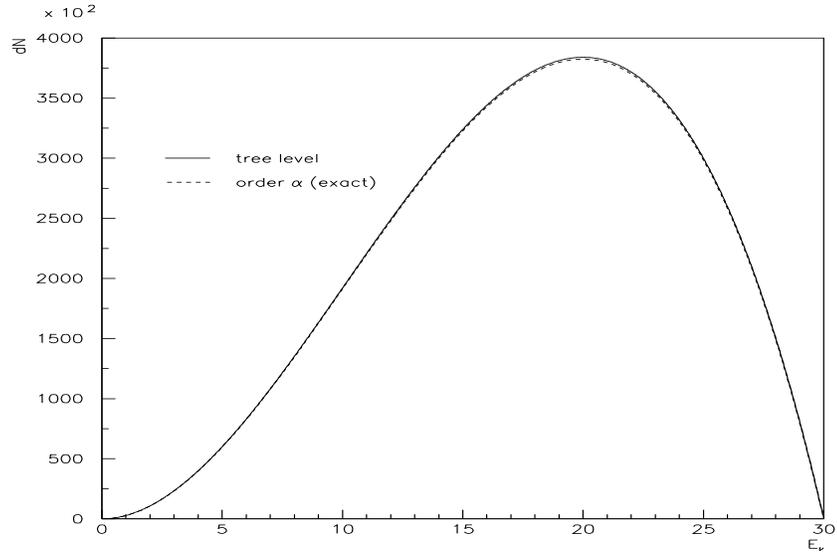, height=8cm,width=12cm}
\caption{\footnotesize Zeroth order and ${\cal O}(\alpha)$ corrected $\bar\nu_e$ forward distributions. Parent muon parameters as in Fig.~\ref{numu}.}
\label{nue}
\end{figure}

Fig.~\ref{numu} and Fig.~\ref{nue} compare the corrected and the tree
level forward $\nu_\mu$ and $\nu_e$ distributions, respectively. In both cases, the relative correction is of ${\cal O}(0.1\%)$, well below the order of the expected precision in the knowledge of the beam parameters. 

The correction found of ${\cal O}(\frac{\alpha}{\pi})\sim {\cal
  O}(0.1\%)$ agrees of that expected from first order QED
  processes. This result differs with the correction of ${\cal
  O}(1\%)$ found for the electron  distribution \cite{Beherends}.
The enhacement of the correction the $e^-$ case is due to the
  ``leading logs'': terms
  proportional to $\ln(\frac{m_\mu}{m_e})$ which stem from the
  emission of collinear photons in the electron leg \cite{LL}.
  Since neutrinos
  are not sensitive to QED, no term in
$\frac{\alpha}{\pi}\ln(\frac {m_\mu}{m_\nu})$ appears in the neutrino
  distributions and, neither, terms in  $\frac{\alpha}{\pi}\log (\frac {m_\mu}{m_e})$,  which cancel when the variables affected by QED corrections, i.e.
electron and the photon electron momenta, are integrated over. An
  identical cancellation it is found in the ${\cal O}({\alpha})$
  corrections to the muon lifetime computed which result to be of
${\cal O}(\frac{\alpha}{\pi})\sim {\cal O}(0.1\%)$ \cite{Beherends}. 

In the laboratory frame, neutrino fluxes are boosted along the muon momentum direction. The formulae of the corrected distributions 
in that frame are given in Appendix B. 

\section{Muon-beam divergence}

We study below the systematic uncertainty in the neutrino distributions produced by the muon beam divergence. For the sake of illustration, the 
quantitative results will be given for a $30$ GeV unpolarized muon beam decaying in a long straight section pointing to a far detector 
located at $2810$ km.

The natural decay angle of the forward neutrino beam in the laboratory frame is deduced from the relation 
between the rest and laboratory frames. In the rest frame, half of the neutrinos are 
emitted within the cone $\theta\leq\pi/2$. In the laboratory frame:
\bea
\cos \theta^{'} = \frac{\cos\theta +\beta}{1 +\beta\cos\theta},
\eea 
where  $ \beta=\sqrt{1-\gamma^{-2}}$ is the muon velocity in the laboratory frame. Therefore, half of neutrinos are emitted 
within the cone subtended by the decay angle $\theta^{'} \leq 1/\gamma$. For instance, for $30$ GeV muons $1/\gamma=m_\mu/E_\mu=3$ mrad.   

\begin{figure}[ht]
\centering
\epsfig{file=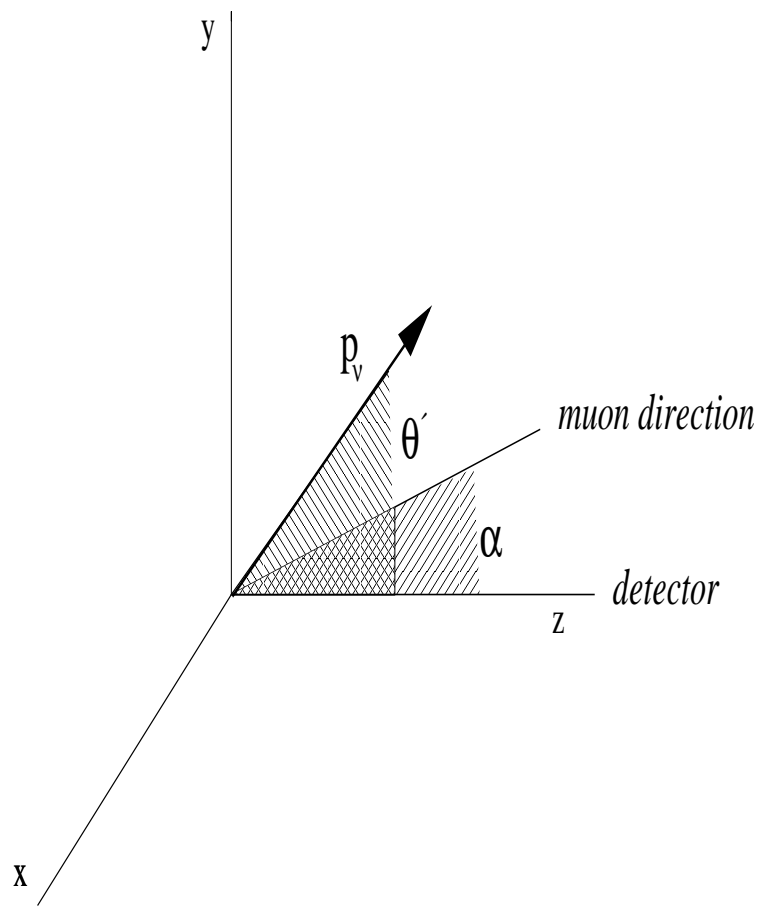,height=12cm, width=10 cm} 
\caption{\footnotesize Muon divergence in the laboratory frame}
\label{divergence}
\end{figure}

For the beam and baseline illustrated here, a $10$ kt detector and one year of data taking \cite{new}, 
 the statistical error on the neutrino flux 
 is of the order of  O($0.4\%$). It is then convenient to restrain the uncertainty induced by the muon beam divergence below that level.
To achieve this, the direction of the beam must be carefully monitored within the decay straight section by placing beam position 
monitors at its ends.  The angular divergence of the parent muon beam is then small compared to the natural decay angle of the neutrino 
beam $\theta^{'}\sim 1/\gamma$, see Fig.~\ref{divergence}, aiming at present to a divergence of  ${\cal O}(0.1 /\gamma)$. 
It implies that the neutrino beam will be collinear, within the limits set by the decay kinematics.

In our calculations we parameterize this beam  focalization by a gaussian distribution with standard deviation $\sigma\sim 0.1/\gamma$ 
(i.e. $0.3$ mrad for $30$ GeV muon beam) \cite{geer}, which suppresses the flux of neutrinos as they separate from the straight direction.
The divergence is introduced analytically by considering that the muon direction opens an angle $\alpha$ with respect to the z-axis, defined 
as the direction pointing towards the far detector at a distance $L$, see Fig.~\ref{divergence}. The neutrino distributions in the rest frame, Eq.~(\ref{treedistr}), are Lorentz boosted along the z-axis. 
The rest-frame basis $(x,\cos{\theta})$ is 
transformed to the lab-frame basis $(z, \cos{\theta}^{'})$, where $z=E_{\nu}/E_{\mu}$ and $\theta^{'}$ is the angle between the neutrino 
beam and the z-axis.
Using the parameters $\beta=\sqrt{1-\gamma^{-2}}$, the boosted distributions read:
\bea
\frac{d^2 N_{\bar\nu_\mu, \nu_\mu}}{dz d\Omega} &=& 
   \frac{4 n_\mu}{\pi L^2 m_\mu^6} \, E_\mu^4 z^2 (1 - \beta (\sin \varphi^{'} \sin \alpha \sin \theta^{'} + \cos \alpha \cos \theta^{'}))\nn\\
 & &   \times \left\{\left[3 m_\mu^2 - 4 z E_\mu^2 (1-\beta (\sin \varphi^{'} \sin \alpha \sin \theta^{'} + \cos \alpha \cos \theta^{'}))\right] 
\right. \nn \\\nn \\
    & & \left. \mp \, {\cal P}_\mu 
   \left[m_\mu^2 - 4 z (1 - \beta(\sin \varphi^{'} \sin \alpha \sin \theta^{'} + \cos \alpha \cos \theta^{'}))\right] \right\},\, \nn \\\nn \\
\frac{d^2 N_{\nu_e, \bar\nu_e}}{dz d\Omega} &=& 
   \frac{24 n_\mu}{\pi L^2 m_\mu^6} \, E_\mu^4 z^2 (1 - \beta(\sin \varphi^{'} \sin \alpha \sin \theta^{'} + \cos \alpha \cos \theta^{'}))\nn\\
& &   \times \left\{ \left[ m_\mu^2 - 2 z E_\mu^2 (1-\beta(\sin \varphi^{'} \sin \alpha \sin \theta^{'} + \cos \alpha \cos \theta^{'}))\right] \right. \nn \\\nn \\
   & & \left. \mp \, {\cal P}_\mu 
   \left[m_\mu^2 - 4 z (1 - \beta(\sin \varphi^{'} \sin \alpha \sin \theta^{'} + \cos \alpha \cos \theta^{'}))\right] \right\}.
\eea  

The above expressions are integrated on $\alpha$, weighted with the gaussian factor
\be 
\frac{e^\frac{-\alpha^{2}}{2\sigma^{2}}}{\sqrt{2\pi \sigma^{2}}}.
\ee

For unpolarized muons ($P_{\mu}=0$) (for different muon polarizations we obtain similar results), it results:
\bea
\nn\\
\frac{d^2 N_{\bar\nu_\mu, \nu_\mu}}{dz d\Omega} &=& 
   \frac{4 n_\mu}{\pi L^2 m_\mu^6} \, E_\mu^4 z^2 \, \left\{3 m_\mu^2 \left( 1 -\beta e^{\frac{-\sigma^{2}}{2}} \cos \theta^{'} \right) \right.\nn\\
 & &\left.- 4 z E_\mu^2 \left(1- 2\beta e^{\frac{-\sigma^{2}}{2}} \cos \theta^{'} 
 \right.\right.\nn \\ 
 & &\left.\left.+\beta^{2}\left(\frac{1 - e^{-2 \sigma^{2}}}{2} \right)\sin^{2}\theta^{'} \sin^{2}\varphi^{'} +\beta^{2}\left(\frac{1 + e^{-2 \sigma^{2}}}{2} \right)\cos^{2}\theta^{'} \right)\right\},\nn\\\nn\\\nn\\\nn\\\nn\\
\frac{d^2 N_{\nu_e \bar\nu_e}}{dz d\Omega} &=& 
   \frac{24 n_\mu}{\pi L^2 m_\mu^6} \, E_\mu^4 z^2 \, \left\{m_\mu^2 \left( 1 -\beta e^{\frac{-\sigma^{2}}{2}} \cos \theta^{'} \right) \right.\nn\\
& &\left.- 2 z E_\mu^2 \left(1- 2\beta e^{\frac{-\sigma^{2}}{2}} \cos \theta^{'}   \right.\right.\nn \\ 
 & &\left.\left.+\beta^{2}\left(\frac{1 - e^{-2 \sigma^{2}}}{2} \right)\sin^{2}\theta^{'} \sin^{2}\varphi^{'}+\beta^{2}\left(\frac{1 + e^{-2 \sigma^{2}}}{2} \right)\cos^{2}\theta^{'} \right)\right\}.
\eea 
Setting $\theta^{'}=0$, the expression of forward neutrino fluxes reads:
\bea
\frac{d^2 N_{\bar\nu_\mu, \nu_\mu}}{dz d\Omega} &=& 
   \frac{4 n_\mu}{\pi L^2 m_\mu^6} \, E_\mu^4 z^2 \, \left\{3 m_\mu^2 \left( 1 -\beta e^{\frac{-\sigma^{2}}{2}} \right) \right.\nn\\
 & &\left.- 4 z E_\mu^2 \left(1- 2\beta e^{\frac{-\sigma^{2}}{2}} 
 +\beta^{2}\left(\frac{1 + e^{-2 \sigma^{2}}}{2} \right)\right)\right\},\nn\\\nn \\
\frac{d^2 N_{\nu_e \bar\nu_e}}{dz d\Omega} &=& 
   \frac{24 n_\mu}{\pi L^2 m_\mu^6} \, E_\mu^4 z^2 \, \left\{m_\mu^2 \left( 1 -\beta e^{\frac{-\sigma^{2}}{2}} \right) \right.\nn\\
& &\left.- 2 z E_\mu^2 \left(1- 2\beta e^{\frac{-\sigma^{2}}{2}} +\beta^{2}\left(\frac{1 + e^{-2 \sigma^{2}}}{2} \right)\right)\right\}.
\eea  
 
\begin{figure}[t]
\centering
\epsfig{file=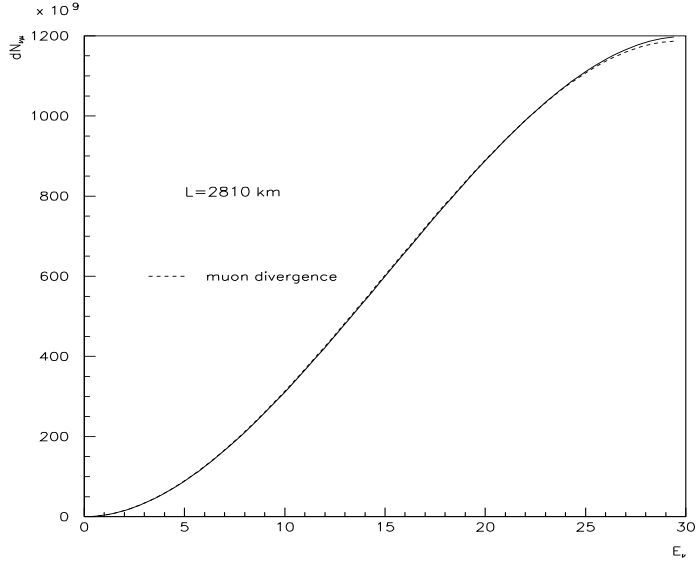, height=8cm,width=10cm}
\caption{\footnotesize $\nu_\mu$ and $\bar\nu_\mu$ differential distributions. The solid lines represent the spectra obtained by averaging over an angular divergence of $0.1$ mrad and the dashed lines the spectra including muon beam divergence. The distributions are plotted in the forward direction $\cos\theta=0$ pointing towards a detector located 2810 km from a the neutrino source
of unpolarized positive or negative muons circulating in the storage ring with energies of $30$ GeV.}
\label{spectrum_mu}
\end{figure}

\begin{figure}[t]
\centering
\epsfig{file=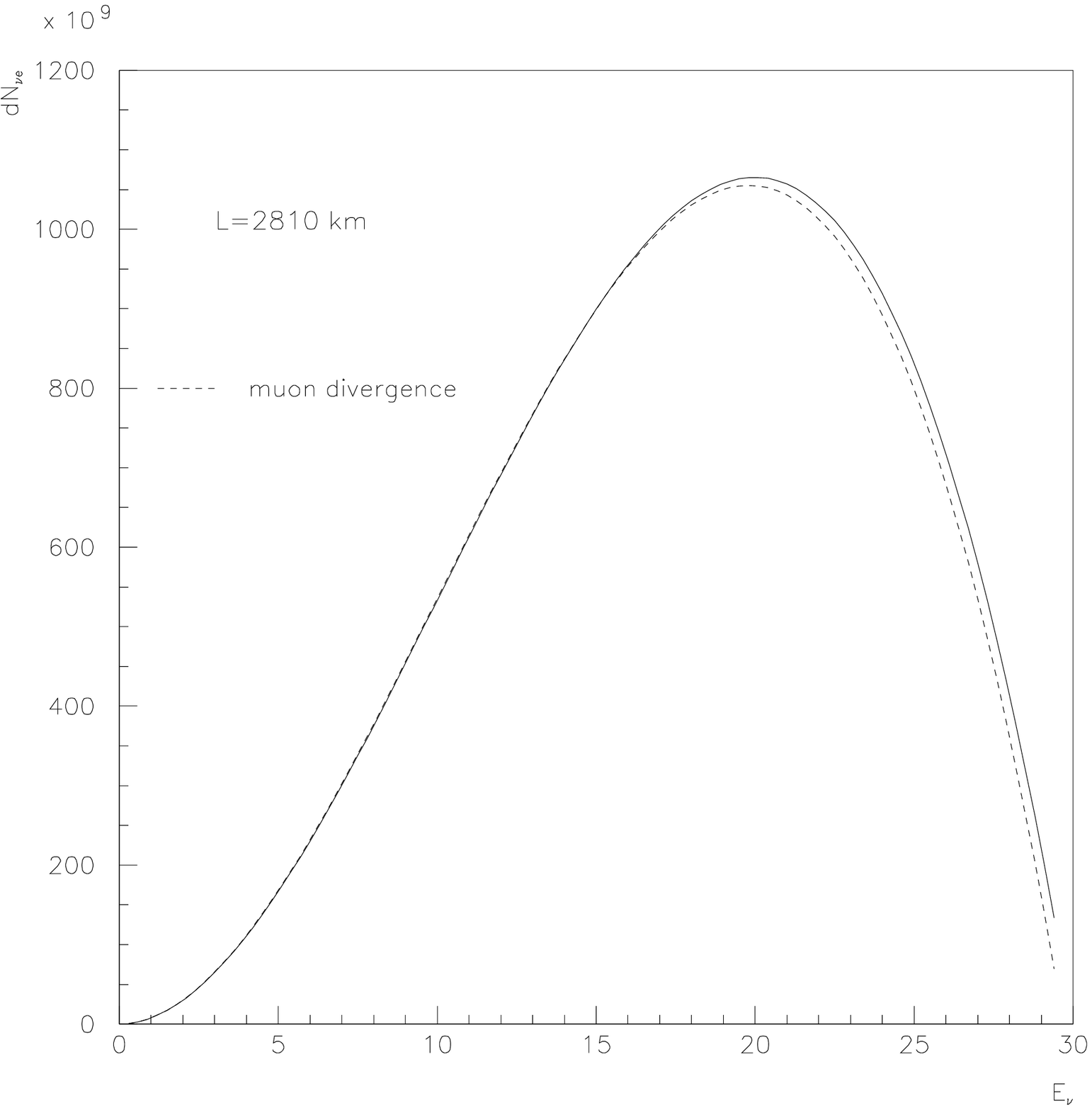, height=8cm,width=10cm}
\caption{\footnotesize $\nu_e$ and $\bar\nu_e$ differential distributions. The solid lines represent the spectra obtained by averaging over an angular divergence of $0.1$ mrad and the dashed lines the spectra including muon beam divergence. The distributions are plotted with the same parameters as of fig(\ref{spectrum_mu}).}
\label{spectrum_e}
\end{figure}

Figs.~\ref{spectrum_mu},~\ref{spectrum_e} show the numerical results for the neutrino and antineutrino spectra in a medium baseline ($2810$ km). We compare 
the distribution where the muon beam is aligned with the detector direction (no beam divergence) with the distribution where the muon-beam divergence is included. In the former, 
neutrino beams are averaged over an angle $\theta^{'}$ of $0.1$ mrad at the far detector \cite{golden}. 

Our formulae predict a similar flux correction 
than previous numerical estimations \cite{geer}. For instance, a $10\%$ uncertainty in the muon beam divergence 
would lead to a flux uncertainty of $0.3\%$.  We obtain,
\bea
\frac{\frac{\Delta d N_{\nu}}{d E}}{\frac{d N_{\nu}}{d E}}\sim 0.03 \frac{\Delta \alpha}{\alpha}.
\eea   
If the muon beam divergence is constrained by lattice design to be less than $0.05/\gamma$, the loss of flux will be negligible \cite{griego}.

\section{Conclusions}

A neutrino factory from muon decay aims at a precision  better than ${\cal O}(1\%)$ in the knowledge of the resulting 
intense neutrino fluxes. 

We have presented here novel results about the effects of QED corrections
and muon beam divergence on the neutrino differential distributions from
muon decay.  We have given the corresponding corrected formulae (for
$m_e=0$ and $m_\nu=0$), including muon polarization effects.  The induced
uncertainties on the neutrino spectra turn out to be a safe ${\cal O}(0.1
\%)$.

Neutrino one-loop corrected distributions diverge at
the upper edge of the kinematical allowed region. This results from a
failure in the cancellation of infrared divergences from virtual photons
by real photons. Applying the soft photon limit to the exact calculations,
we have isolated the end-point divergent term for the neutrino
distributions which takes the form of $\ln^2(1-x)$. In order to control
this singularity, the double logarithmic-contribution is
exponentiated, encompassing the contributions from all orders of
perturbation theory. All in all, the exact neutrino distributions get
corrections of ${\cal O}(0.1)\%$, safely below the expected precision in
the flux measurements.

We have also studied carefully the influence of the muon beam divergence on the neutrino spectra at the far site. 
The challenge in designing the neutrino production section, where the muons decay, is to constrain the muon beam divergence 
to a value smaller than the natural cone of forward going neutrinos in the laboratory frame, ($\sim 1/\gamma$).  
At present, the long straight sections under discussion aim at an angular muon beam divergence of the order of $0.1/\gamma$, typically less than one mrad. 

\section{Acknowledgments}

We thank M.B. Gavela, P. Hern\'andez and A. De R\'ujula for their physics suggestions and useful discussions. We thank as well F.J Yndur\'ain for illuminating conversations. We are further indebted to  A. Blondel, F. Dydak, J.J.~G\'omez-C\'adenas. A.B acknowledges M.E.C.D for financial support by FPU grant AP2001-0521 and O.M acknowledges C.A.M for financial support by a FPI grant. The work has been partially supported by CICYT FPA2000-0980 project.

\newpage

{\Large{\bf Appendix A: QED loop corrections}}
\vspace{0.4cm}

{\bf A.1 Virtual corrections}

\begin{figure}
\centering
\epsfig{file=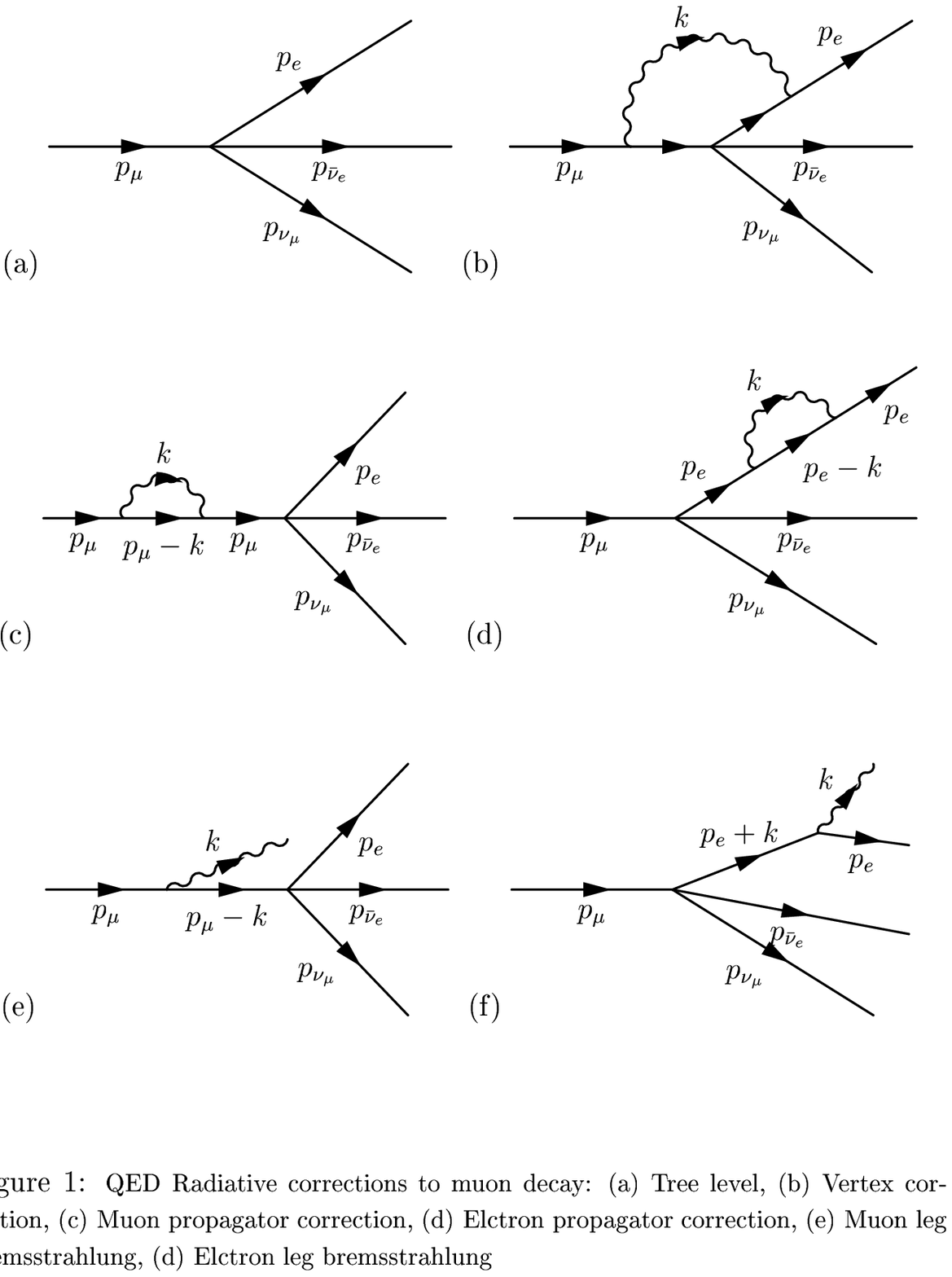}
\caption{\footnotesize QED Radiative corrections to muon decay: (a) Tree level diagram, (b) Vertex correction, (c) Muon propagator correction, (d) Electron propagator correction, (e) Muon leg bremsstrahlung, (d) Electron leg bremsstrahlung}
\label{loops}
\end{figure}

There three diagrams containing a photon loop: the exchange of the virtual photon between the muon and the 
electron legs, Fig.~\ref{loops}b, and lepton propagator corrections, Figs.~\ref{loops}c,~\ref{loops}d. They correct the invariant amplitude of 
the muon decay as follows:
\be
-i\ M = \frac{G_F}{\sqrt{2}}\{\bar{u}(p_e)\Gamma_\sigma {u}(p_\mu) \}
                              \{\bar{u}(p_{\nu_\mu})\gamma^\sigma(1-\gamma_5) {v}(p_{\bar\nu_e}) \}~.
\ee
$\Gamma_\sigma$ is the corrected $\mu-e$ vertex:
\be
\label{gam_sig}
\Gamma_\sigma= \gamma_\sigma(1-\gamma_5)+ \Gamma_\sigma^b+\Gamma_\sigma^{c,d}~,
\ee
where $\Gamma_\sigma^b$ results from the diagram in Fig.~\ref{loops}b and $\Gamma_\sigma^{c,d}$ from those  
in Figs.~\ref{loops}c,~\ref{loops}d.

After integration over the photon momentum, the correction from
diagram ~\ref{loops}b, has the expression
\bea
\label{V}
\Gamma_\sigma^b  =   - \frac{\alpha}{2\pi}\,&[& 
            (g_{\rm{IR}}^{b}+g_{\rm{UV}}^{b})\,\gamma_\sigma\,(1-\gamma_5)
            +g_{\rm{R}}^{\rm{S}}\,\gamma_\sigma\,(1+\gamma_5)\nn \\
            &+& g_{\rm{L}}^{\rm{V}}\,p_{1\sigma}\,(1-\gamma_5)
            + g_{\rm{R}}^{\rm{V}}\,p_{2\sigma}\,(1+\gamma_5)\,]~,
\eea
whith
\bea
\label{IRvert} 
g_{\rm{IR}}^{b} & = &  \coth{\phi} \left  [   
          {\rm L}\left( \frac {2\sinh{\phi}} {e^{\omega}-e^{-\phi}} \right)            - {\rm L}\left( \frac {2\sinh{\phi}} {e^{\phi}-e^{-\omega}} \right)            + (\omega-\phi)
            \ln{\left( \frac {2\sinh {\left( \frac{\omega-\phi}{2} \right)}}                              {2\sinh {\left( \frac{\phi+\omega}{2} \right)}}                 \right)}
                    \right.  \nn \\  
        &\,& \qquad  \left. \,\,+\,\phi\left(\omega
                  -\ln{ \left( \frac{\lambda^2}{m_e^2} \right) } \right)
                      \right]~, \nn\\
\label{UVvert} 
g_{\rm{UV}}^{b}  & = & \frac {\phi\sinh{\phi}-\omega\sinh{\omega}}
                      {2(\cosh{\omega}-\cosh{\phi})} + \frac{1}{2}\left(\omega                      -  \ln{ \left( \frac{\Lambda^2}{m_e^2} \right) } \right)
                     -  \frac{3}{2}, \nn \\
g_{\rm{R}}^{\rm{S}}& = &\frac{- \phi}{\sinh{\phi}},\nn\\
g_{\rm{L}}^{\rm{V}}& = & \frac{1}{2m_{\mu}\sinh{\phi}} \left[ \phi - 
            \frac {\omega\sinh{\phi}-\phi\sinh{\omega}}
                  {\cosh{\omega}-\cosh{\phi}} \right],\nn\\
g_{\rm{R}}^{\rm{V}}& = & \frac{1}{2m_{e}\sinh{\phi}} \left[ \phi + 
            \frac {\omega\sinh{\phi}-\phi\sinh{\omega}}
                  {\cosh{\omega}-\cosh{\phi}} \right]~,
\eea

where the IR term is regularized by a finite photon mass $\lambda$ and
the variables
\be
\cosh{\phi}=\frac{(p_\mu p_e)}{m_{\mu}m_e}\qquad
e^{\omega}=\frac{m_\mu}{m_e}~.
\ee
are introduced following \cite{Beherends}.

L(x) is the Spence function 
\be
L(x)\equiv -\int_0^x\,dt\,\frac{\ln|1-t|}{t}~.
\ee

The contribution of self-energy diagrams to the muon-electron vertex,
after integration over the photon momentum, is given by
\be
\Gamma^{c,d}_\sigma= 
       -\frac{\alpha}{2\pi}\,\frac{1}{2}\,(h_{\rm{UV}}^{c,d}+h_{\rm{IR}}^{c,d})~,
\ee
where, now,
\bea
\label{UVself}
h_{\rm{UV}}^{c,d} & = & - \frac{1}{2}\left(\omega -
                    \ln{ \left( \frac{\Lambda^2}{m_e^2} \right) } \right)
                 +  \frac{3}{2}~,\\
\label{IRself}
h_{\rm{IR}}^{c,d} & = &  \left(\omega
                    - \ln{ \left( \frac{\lambda^2}{m_e^2} \right) } \right)+2~. 
\eea

Adding Eq.~(\ref{UVvert}) and Eq.~(\ref{UVself}) the UV divergences
are exactly cancelled. The IR terms in Eq.~(\ref{IRself}), when combined with Eq.~(\ref{V}) gives rise to the term 
\bea
\label{gls}
g_{\rm{L}}^{\rm{S}} & = & - \coth{\phi} \left[ 
        \phi  
      - {\rm L}\left( \frac {2\sinh{\phi}} {e^{\omega}-e^{-\phi}} \right)            + {\rm L}\left( \frac {2\sinh{\phi}} {e^{\phi}-e^{-\omega}} \right)            - (\omega-\phi)
        \ln{\left( \frac {2\sinh {\left( \frac{\omega-\phi}{2} \right)}}                              {2\sinh {\left( \frac{\phi+\omega}{2} \right)}}                 \right)}
                       \right] \nn, \\ 
     & + & \frac {\phi\sinh{\phi}-\omega\sinh{\omega}}
                 {2(\cosh{\omega}-\cosh{\phi})} + 2  
                -(1-\phi\coth{\phi}) \left(\omega
                  -\ln{ \left( \frac{\lambda^2}{m_e^2} \right) } \right)~.
\eea

\vspace{0.5cm}
{\bf A.2 Bremstrahlung corrections}

The contribution from real photon emission, Fig.~\ref{loops}e and Fig.~\ref{loops}f, is given by 
\be
d\Gamma_r=\frac{1}{2m_{\mu}}\,64\,{\rm G_F}^2\,\,|M_R|^2\,d\Phi_4(p_\mu;p_e,p_{\bar\nu_e},p_{\nu_\mu},k),
\ee
where the amplitude $|M_R|^2$ has the following expression:
\be
\label{Mreal}
|M_R|^2= \frac{\alpha}{2\pi}\,\bigg[\frac{A}{(p_\mu
 k)^2}+\frac{B}{(p_e k)^2}-\frac{C}{(p_\mu k)(p_e k)}\bigg].
\ee

The numerators for unpolarized muons read:
\bea
\label{ampl_real}
A & = & p_\mu^2 \,[(p_\mu p_{\bar\nu_e})(p_e p_{\nu_\mu})-(kp_{\bar\nu_e})(p_e p_{\nu_\mu})-(p_\mu k)(kp_{\bar\nu_e})(p_e p_{\nu_\mu})] \nn \\ 
B & = & p_e^2 \,[(p_\mu p_{\bar\nu_e})(p_e p_{\nu_\mu})+(p_\mu p_{\bar\nu_e})(kp_{\nu_\mu})-(p_\mu p_{\bar\nu_e})(p_e k)(kp_{\nu_\mu})] \nn \\ 
C & = & (p_\mu p_e)\,[2(p_\mu p_{\bar\nu_e})(p_e p_{\nu_\mu})+(p_\mu p_{\bar\nu_e})(kp_{\nu_\mu})-(kp_{\bar\nu_e})(p_e p_{\nu_\mu})]  \\      & 
+ & (p_e k)[(p_\mu p_{\bar\nu_e})(p_e p_{\nu_\mu})+(p_\mu p_{\bar\nu_e})(p_\mu p_{\nu_\mu})-(p_\mu p_{\bar\nu_e})(kp_{\nu_\mu})]  \nn \\         & - & (p_\mu k)[(p_\mu p_{\bar\nu_e})(p_e p_{\nu_\mu})+(p_e p_{\bar\nu_e})(p_e p_{\nu_\mu})-(kp_{\bar\nu_e})(p_e p_{\nu_\mu})].\nn
\eea
Terms of order $k^2$  are not included in Eq.~(\ref{ampl_real}), since they vanish in the limit of massless photons. 

\newpage
{\Large{\bf Appendix B: QED corrected distributions
\vspace{0.1cm}
in the laboratory frame}}
\vspace{0.3cm}

In order to obtain the neutrino distributions in the laboratory frame, a Lorentz boost is performed in the direction of the muon velocity towards the detector at a distance $L$. The rest-frame basis $(x,\cos{\theta})$ is transformed to the lab-frame basis $(z, \cos{\theta}^{'})$, where $z=E_{\nu}/E_{\mu}$ is the scaled energy at the laboratory frame and $\theta^{'}$ is the angle between the neutrino beam and the direction of the muon beam \cite{golden}. The muon beam divergence is set to zero. Using the parameters $\gamma=E_\mu/m_\mu$ and $\beta=\sqrt{1-\gamma^{-2}}$, the boosted distributions read:
\bea
\frac {d^2 N_{\nu_{\mu}} }{ dz d\cos{\theta}^{'} }& = &  \,F_{\nu_\mu}^{(0)} (z,\theta^{'}) + {\cal P}_\mu\, J_{\nu_\mu}^{(0)}(z,\theta^{'}) \cos \theta^{'}  \nn\\\nn\\
&- &\frac{\alpha}{2\pi}\left[ F_{\nu_{\mu}}^{(1)}(z,\theta^{'})
+ {\cal P}_\mu\,  J_{\nu_\mu}^{(1)}(z,\theta^{'}) \cos \theta^{'}\right]\,,\nn \\ \\
\frac {d^2 N_{\bar\nu_e} }{ dz d\cos{\theta}^{'}} & = & \,F_{\bar\nu_e}^{(0)}(z,\theta^{'}) + {\cal P}_\mu\, J_{\bar\nu_e}^{(0)}(z,\theta^{'}) \cos \theta^{'}\nn \\\nn\\
&- &\frac{\alpha}{2\pi}\left[ F_{\bar\nu_e}^{(1)}(z,\theta^{'})
+ {\cal P}_\mu\, J_{\bar\nu_e}^{(1)}(z,\theta^{'})\cos \theta^{'}\right]\,, 
\eea
where
\bea
F_{\nu_{\mu}}^{(0)}(z,\theta^{'}) & = & 8 \, \frac{E_\mu^4}{m_\mu^6}\,  z^2\, (1-\beta\cos{\theta}^{'})(3m_{\mu}^2-4E_{\mu}^2 z(1-\beta\cos{\theta}^{'})),\\ 
\nn \\ 
J_{\nu_{\mu}}^{(0)}(z,\theta^{'}) & = & 8  \, \frac{E_{\mu}^4}{m_\mu^6}\, z^2\, (1-\beta\cos{\theta}^{'})(m_{\mu}^2-4E_{\mu}^2 z(1-\beta\cos{\theta}^{'})), \\ \nn  \\
F_{\bar\nu_e}^{(0)}(z,\theta^{'}) & = & 48 \frac{E_{\mu}^4}{m_{\mu}^6}\,z^2\,                             (1-\beta\cos{\theta}^{'})(m_{\mu}^2-2E_{\mu}^2z(1-\beta\cos{\theta}^{'})),\\ \nn \\
J_{\bar\nu_e}^{(0)}(z,\theta^{'}) & = & 48 \frac{E_{\mu}^4}{m_{\mu}^6}\,z^2\,                                (1-\beta\cos{\theta}^{'})(m_{\mu}^2-2E_{\mu}^2z(1-\beta\cos{\theta}^{'})),\\ \nn \\ 
F_{\nu_{\mu}}^{(1)}(z,\theta^{'})  & = & 
        F_{\nu_\mu}^{(0)}(z,\theta^{'}) k(z,\theta^{'})
        +\frac{1}{3\J}\Bigg\{\left[41-36\x \right. \nn \\
        & + & \left.42\x^2-16\x^3\right]\nn \\
        & \times & \ln{\xm} \\
        & + & \frac{1}{2}\x\left[82-153\x \right. \nn  \\
        & + & \left. 86\x^2\right] \Bigg\},\nn \\ \nn \\\nn \\ \nn \\
J_{\nu_{\mu}}^{(1)}(z,\theta^{'})  & = & 
        J_{\nu_\mu}^{(0)}(z,\theta^{'}) k(z,\theta^{'})
        +\frac{1}{3\J}\Bigg\{\left[11- 36\x \right. \nn \\
        & + & \left.14\x^2-16\x^3 \right. \nn \\
        & + & \left.4\x^{-1} \right]       
        \times  \ln{\xm} \\
        & + & \frac{1}{2}\left[-8+18\x \right. \nn \\
        & - & \left. 103\x^2 + 78\x^3 \right] \Bigg\},\nn \\ \nn \\\
F_{\bar\nu_e}^{(1)}(z,\theta^{'})  & = & 
        F_{\nu_e}^{(0)}(z,\theta^{'}) k(z,\theta^{'}) 
          +  \frac{2\xm}{\J} \nn\\
        & \times &\Bigg\{\left[5+ 8\x \right. \nn \\
        & + & \left. 8\x^2\right]\ln{\xm}\\
        & + & \frac{1}{2}\,\x\left[10-19\x\right]
                                       \Bigg\},\nn \\ \nn \\      
J_{\bar\nu_e}^{(1)}(z,\theta^{'}) & = &                            J_{\nu_e}^{(0)}(z,\theta^{'}) k(z,\theta^{'})
        +  \frac{2\xm}{\J}\nn \\
        &\times & \Bigg\{\left[-3+ 12\x + 8\x^2 \right. \nn \\
        & + & \left. 4\x^{-1} \right]\,\ln{\xm} \\
        & + & \frac{1}{2}\,\left[8-2\x - 15\x^2 \right]\Bigg\},\nn \\ \nn \\    %
k(z,\theta^{'}) & = & \ln^2{\xm}+ 2{\rm L}\x  + \frac{2\pi^2}{3}
\eea
\newpage

\newpage

\newpage

\begin{figure}[ht]
\centering
\epsfig{file=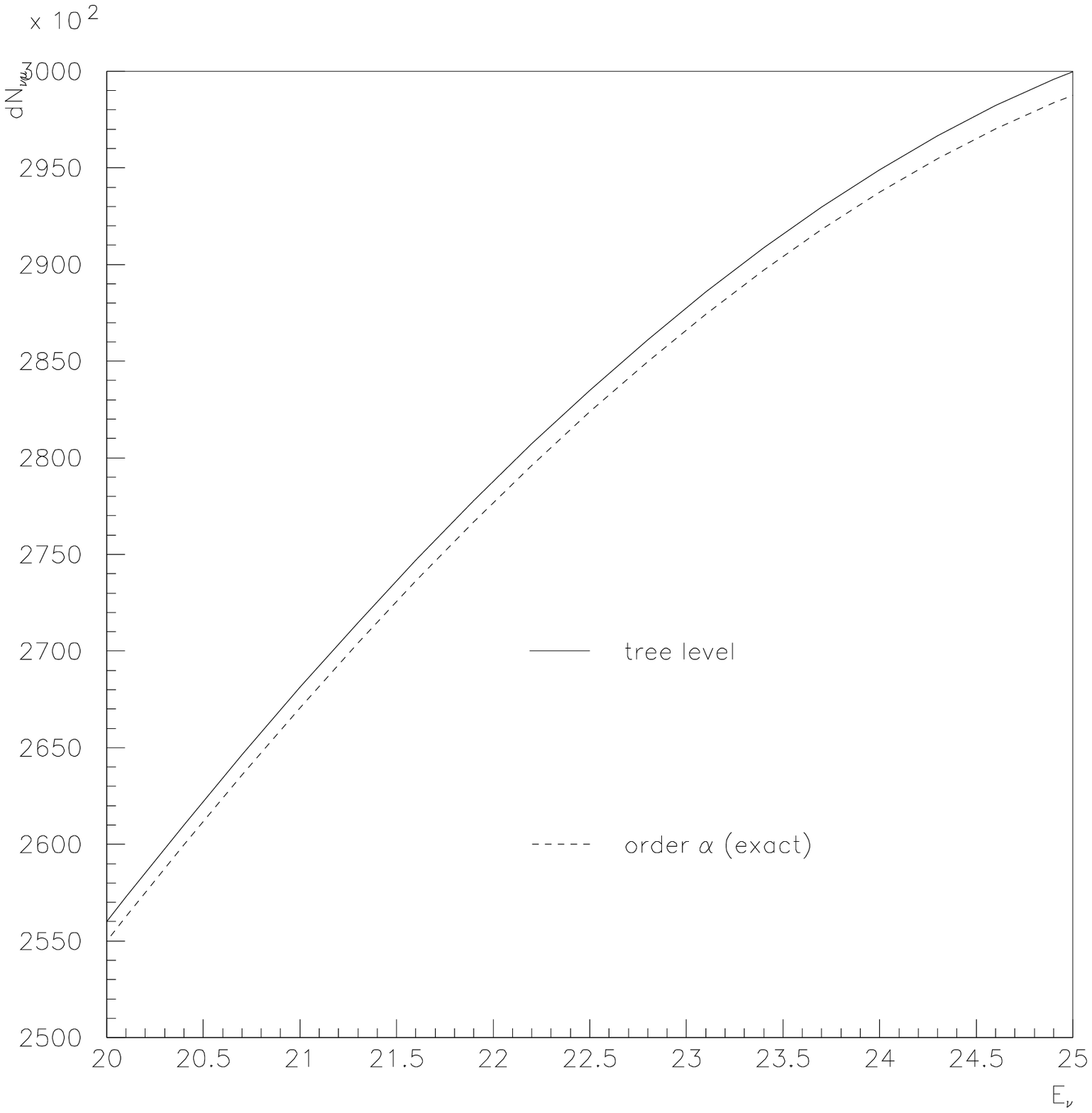, height=8cm,width=12cm}
\caption{\footnotesize Detail of Fig.~\ref{numu}. }
\end{figure}
\begin{figure}[ht]
\centering
\epsfig{file=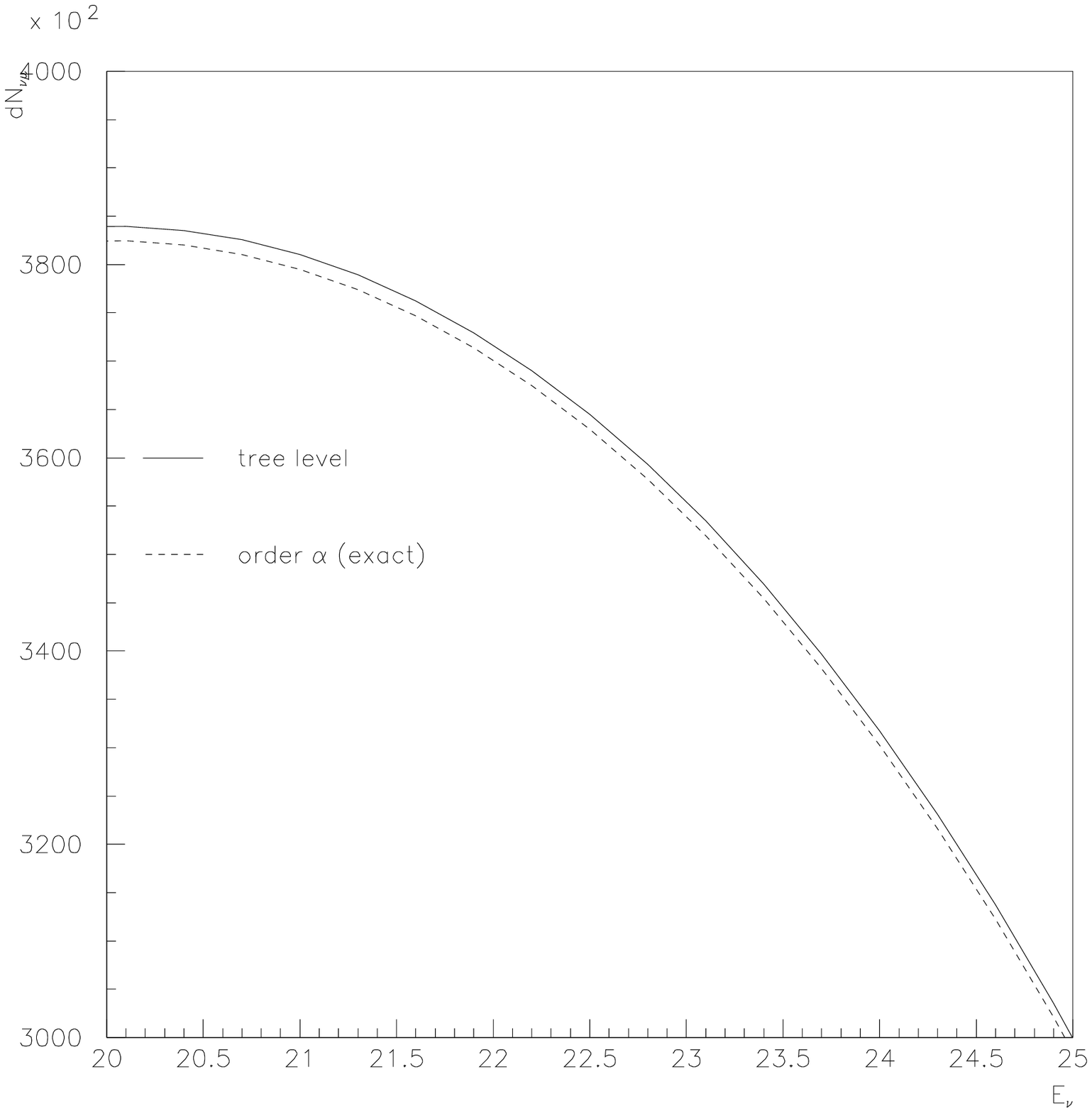, height=8cm,width=12cm}
\caption{\footnotesize Detail of Fig.~\ref{nue}. }
\end{figure}


\begin{thebibliography}{99}
%
\bibitem{SuperK}
Y.~Fukuda {\it et al.}  [SuperKamiokande Collaboration], Phys.\ Rev.\ Lett.\  {\bf 82} (1999) 2644
%
\bibitem{SNO}
J.~N.~Bahcall, P.~I.~Krastev and A.~Y.~Smirnov,
JHEP {\bf 0105} (2001) 015
%
\bibitem{exper}
S.~H.~Ahn {\it et al.}  [K2K Collaboration],
Phys.\ Lett.\ B {\bf 511}, 178 (2001),\\
MINOS Collaboration http://www-numi.fnal.gov:8875/,\\
A.~G.~Cocco  [OPERA Collaboration],
Nucl.\ Phys.\ Proc.\ Suppl.\  {\bf 85}, 125 (2000).
%
\bibitem{Snowmass}
T.~Adams {\it et al.},
in {\it Proc. of the APS/DPF/DPB Summer Study on the Future of Particle Physics (Snowmass 2001) } ed. R.~Davidson and C.~Quigg, arXiv:hep-ph/0111030.
%
\bibitem{Particle} 
D.~E.~Groom {\it et al.}  [Particle Data Group Collaboration],\\
Eur.\ Phys.\ J.\ C {\bf 15} (2000) 1.
%
\bibitem{todos}
V.~D.~Barger {\it et al.}, hep-ph/0103052.
%
\bibitem{pilar}
J.~J.~Gomez-Cadenas {\it et al.}  [CERN working group on Super Beams
                  Collaboration]\\hep-ph/0105297.
\bibitem{Geer1}
S.~Geer,Phys.\ Rev.\ D {\bf 57} (1998) 6989
[Erratum-ibid.\ D {\bf 59} (1998)]
%
\bibitem{Blondel}
A.~Blondel {\it et al.},
Nucl.\ Instrum.\ Meth.\ A  {\bf 451} (2000) 102.
%
\bibitem{Beherends} 
R.J.~Finkelstein, R.~E.~Behrends and A.~Sirlin,
{\em Phys. Rev.} 101 (1956) 866;\\
S.~Berman, {\em Phys. Rev.} 112 (1958) 267;\\
T.~Kinoshita and A.~Sirlin, {\em Phys. Rev.} 113 (1959) 1652.
%
\bibitem{Greub}
C.~Greub, D.~Wyler and W.~Fetscher,
Phys.\ Lett.\ B {\bf 324} (1994) 109
[Erratum-ibid.\ B {\bf 329} (1994) 526]
%
\bibitem{Jezabek1}
M.~Je{\.z}abek and J.~H. K{\"u}hn,
Nucl. Phys. {\bf B320} (1989) 20.
%
\bibitem{Jezabek2}
A.~Czarnecki, M.~Je{\.z}abek and J.H. K{\"u}hn,
Nucl. Phys.{\bf B351} (1991) 70.
%
\bibitem{Jezabek3}
A.~Czarnecki, M.~Je{\.z}abek,
Nucl. Phys. {\bf B427} (1994) 3
%
\bibitem{golden} 
A.~Cervera, A.~Donini, M.~B.~Gavela, J.~J.~Gomez Cadenas, P.~Hernandez, O.~Mena and S.~Rigolin,
Nucl.\ Phys.\ B {\bf 579} (2000) 17, 
[Erratum-ibid.\ B {\bf 593} (2000) 731]
%
%
\bibitem{gaisser} 
T.~K.~Gaisser, {\em ``Cosmic Rays and Particle Physics''}, 
Cambridge University Press, 1990.
%
\bibitem{Matson}
L.~Matsson, Nucl.Phys. {\bf B12} (1969) 647;\\
D.A.~Ross, Nuovo Cim. {\bf 10A} (1972) 475;\\
A.M.~Sachs and A.Sirlin, {\em ``Muon physics''}, ed.V.H.Hughes,1975.
%
\bibitem{Yennie}
D.~Yennie, S.~Frautschi and H.Suura, Ann.Phys. {\bf 13} (1961) 379;\\
S.~Weinberg, Phys. Rev {\bf 140} (1965) B516.
%
\bibitem{LL}
A.~Arbuzov, A.~Czarnecki and A.~Gaponenko, hep-ph/0202102.
%
\bibitem{new}
S.~Geer, C.~Johnstone and D.~Neuffer, FERMILAB-Pub-99/121.
%
\bibitem{geer}
C.~Crisan and S.~Geer, report FERMILAB-TM-2001.
%
\bibitem{griego}
I.~M.~Papadopoulos,
{Nucl.\ Instrum.\ Meth.\ A} 451, 138 (2000).
%
\end{thebibliography}
\end{document}